\documentclass[twocolumn,preprintnumbers,amsmath,amssymb,superscriptaddress,subeqn,prl]{revtex4-1}

\usepackage{amsmath}
\usepackage{stmaryrd}
\usepackage[pdftex]{graphicx} 
\usepackage{epstopdf} 
\usepackage{url}      
\usepackage{times}    
\usepackage{float}
\usepackage[section]{placeins} 
\usepackage{color} 

\usepackage{array} 
\newcolumntype{L}[1]{>{\raggedright\let\newline\\\arraybackslash\hspace{0pt}}m{#1}}
\newcolumntype{C}[1]{>{\centering\let\newline\\\arraybackslash\hspace{0pt}}m{#1}}
\newcolumntype{R}[1]{>{\raggedleft\let\newline\\\arraybackslash\hspace{0pt}}m{#1}}

\newcommand{\bra}[1]{\langle #1|}
\newcommand{\ket}[1]{|#1\rangle}

\newcommand{\red}[1]{\textcolor{black}{#1}}

\begin{document}
\title{Solid-state optical absorption from optimally-tuned time-dependent\\ range-separated hybrid density functional theory}

\author{Sivan Refaely-Abramson}
\affiliation{Department of Materials and Interfaces, Weizmann Institute of Science, Rehovoth 76100, Israel}
\author{Manish Jain}
\affiliation{Department of Physics, Indian Institute of Science, Bangalore 560 012, India}
\author{Sahar Sharifzadeh}
\affiliation{ Department of Electrical and Computer Engineering and Physics Division of Materials Science and Engineering,
 Boston University, Boston, MA 02215 , USA }
\author{Jeffrey B. Neaton}
\affiliation{Department of Physics, University of California Berkeley, Berkeley, CA 94720 , USA}
\affiliation{Molecular Foundry, Lawrence Berkeley National Laboratory, Berkeley, CA 94720, USA}
\affiliation{ Kavli Energy NanoScience Institute at Berkeley, Berkeley, CA 94720, USA}

\author{Leeor Kronik}
\affiliation{Department of Materials and Interfaces, Weizmann Institute of Science, Rehovoth 76100, Israel}

\date{\today}

\begin{abstract}
We present a framework for obtaining reliable solid-state charge and optical excitations and spectra from optimally-tuned range-separated hybrid density functional theory. \red{The approach, which is fully couched within the formal framework of generalized Kohn-Sham theory}, allows for accurate prediction of exciton binding energies. We demonstrate our approach through \red{first principles} 
calculations of one- and two-particle excitations in pentacene, a molecular semiconducting crystal, where our work is in excellent agreement with experiments and prior computations. We further show that with one adjustable parameter, \red{set to produce an accurate bandgap}, this method accurately predicts band structures and optical spectra of silicon and lithium flouride, prototypical covalent and ionic solids. Our findings indicate that for a broad range of extended bulk systems, this method may provide a computationally inexpensive alternative to many-body perturbation theory, opening the door to studies of materials of increasing size and complexity. 
\end{abstract}

\maketitle

Many solid-state systems exhibit strong excitonic effects, notably an optical excitation spectrum that is affected substantially by interaction between excited electron and hole quasiparticle states. The nature of this electron-hole, or {\it excitonic}, interaction is of central importance for a variety of applications in, e.g., optoelectronics and photovoltaics \cite{Savoie2014}. Nevertheless, its accurate theoretical prediction remains a challenging task. It is common to account for such interactions within the framework of ab initio many-body perturbation theory, in which single-particle excitations are well-predicted from Dyson's equation, typically solved within the GW approximation \cite{Hedin1965,Hybertsen1986}, and two-particle excitations are well-predicted using the Bethe-Salpeter equation (BSE) \cite{Rohlfing1998, *Rohlfing2000, Strinati1982, *Strinati1984}. 

Current GW-BSE calculations are highly demanding and therefore presently impose significant practical limits on the calculated system size and complexity. Density functional theory (DFT), in both its time-independent \cite{DreizlerGross,ParrYang} and time-dependent (TDDFT) \cite{Marques2012,Casida1995,Burke2005,Ullrich_book} forms, is considerably more efficient computationally. However, common (semi-)local approximations to both DFT and TDDFT suffer from serious deficiencies which have precluded their use as a viable alternative to GW-BSE in the prediction of excitonic properties \cite{Onida2002}. First, quasi-hole and quasi-electron excitation energies are generally underestimated and overestimated, respectively, by the DFT Kohn-Sham eigenvalue spectrum \cite{Kummel2008,Kronik2012}. While the same functionals often perform better in the prediction of optical excitation energies of isolated molecular systems, the Kohn-Sham gap is typically similar to the optical gap \cite{Salzner1997,Chong2002,Baerends2013,Kronik2012,Kronik2014}. In any case, they still fail in the solid-state limit \cite{Onida2002,Ullrich_book,Izmaylov2008,Sharma2014,Ullrich2014}. Therefore, neither one- nor two-particle excitations are well-predicted in the solid-state, and hence the nature of excitons or their binding energies are not obtained.

The failure of semi-local functionals in predicting solid-state absorption spectra has been traced back to an incorrect description of the long-range electron-electron and electron-hole interaction, manifested by the absence of a $1/q^2$ contribution \cite{Gonze1995,*Ghosez1997,Kim2002} to the interaction, where $q$ is a wavevector in the periodic system. Several ingenious schemes for overcoming this deficiency have been suggested, including the use of  an exchange-correlation kernel of the form $f_{xc}(r,r')=-\alpha/(4\pi |r-r'|)$, where $\alpha$ is a system-dependent empirical parameter  \cite{Reining2002, Botti2004}; a static approximation to the exchange-correlation kernel based on a jellium-with-gap model \cite{Trevisanutto2013};  a ``bootstrap'' parameter-free kernel, achieved using self-consistent iterations of the random phase approximation (RPA) dielectric function  \cite{Sharma2011,Sharma2014}; a related "guided iteration" RPA-bootstrap kernel \cite{Rigamonti2015}; and the Nanoquanta kernel \cite{Onida2002, Reining2002, Sottile2003, Marini2003, *Adragna2003}, derived by constructing the exchange-correlation kernel from an approximate solution to the BSE. Each correction provides a major step forward. However, none is a fully DFT-based solution, as single quasiparticle excitations are obtained from GW, RPA, a DFT+U approach, or a scissors-shift correction.

A different path for enabling TDDFT calculations in the solid state is the use of (global or range-separated) hybrid functionals. These are still well within density functional theory, using the generalized Kohn-Sham (GKS) framework \cite{Seidl1996,Kummel2008,Kronik2012}, and their non-local Fock-like exchange component assists in the inclusion of long-range contributions. Although the time-dependent GKS equations have yet to be formally derived, hybrid functionals are already widely used for calculating optical properties. For gas-phase molecules, hybrid functionals can improve optical excitation energies, although standard hybrids still do not provide for accurate single-particle excitation energies \cite{Salzner1997,Kummel2008,Kronik2012,Kronik2014}. TDDFT using the Heyd-Scuseria-Ernzerhof (HSE) short-range hybrid functional \cite{Heyd2003, *Heyd2006}, where non-local exchange is introduced only in the short-range, can improve the absorption spectra of semiconductors and insulators \cite{Paier2008}, although some discrepancies remain. However, the HSE functional still does not provide the desired long-range non-local contribution. The B3LYP hybrid functional \cite{Becke1993, *Stephens1994}, in which a global 20\% fraction of exact-exchange is used, was recently shown to yield TDDFT optical spectra for semiconductors in good agreement with experiment \cite{Bernasconi2011, *Tomic2014}, \red{but in some cases a larger fraction of exact exchange was needed \cite{Ferrari2015}}. Although in this case a non-local contribution to the kernel tail does exist, it is global and parameterized for a finite set of small organic molecules. However, global and short-range hybrid functionals were shown to be insufficient predictors of band-structures in solid-state systems \cite{Jain2011}, notably for molecular crystals \cite{Refaely-Abramson2013} where excitonic effects are strong. Recently, Yang et al. \cite{Yang2015} suggested a screened exact-exchange (SXX) approach, in which the local part of the hybrid calculation is set to zero and the time-dependent Hartree-Fock exchange is scaled down non-empirically per system by using the inverse of the dielectric constant, based on a ground state obtained from a scissor-corrected local density approximation (LDA) calculation. Again, this led to improved performance for more strongly bound excitons. 

\begin{figure}
\begin{center}
\includegraphics[scale=0.5]{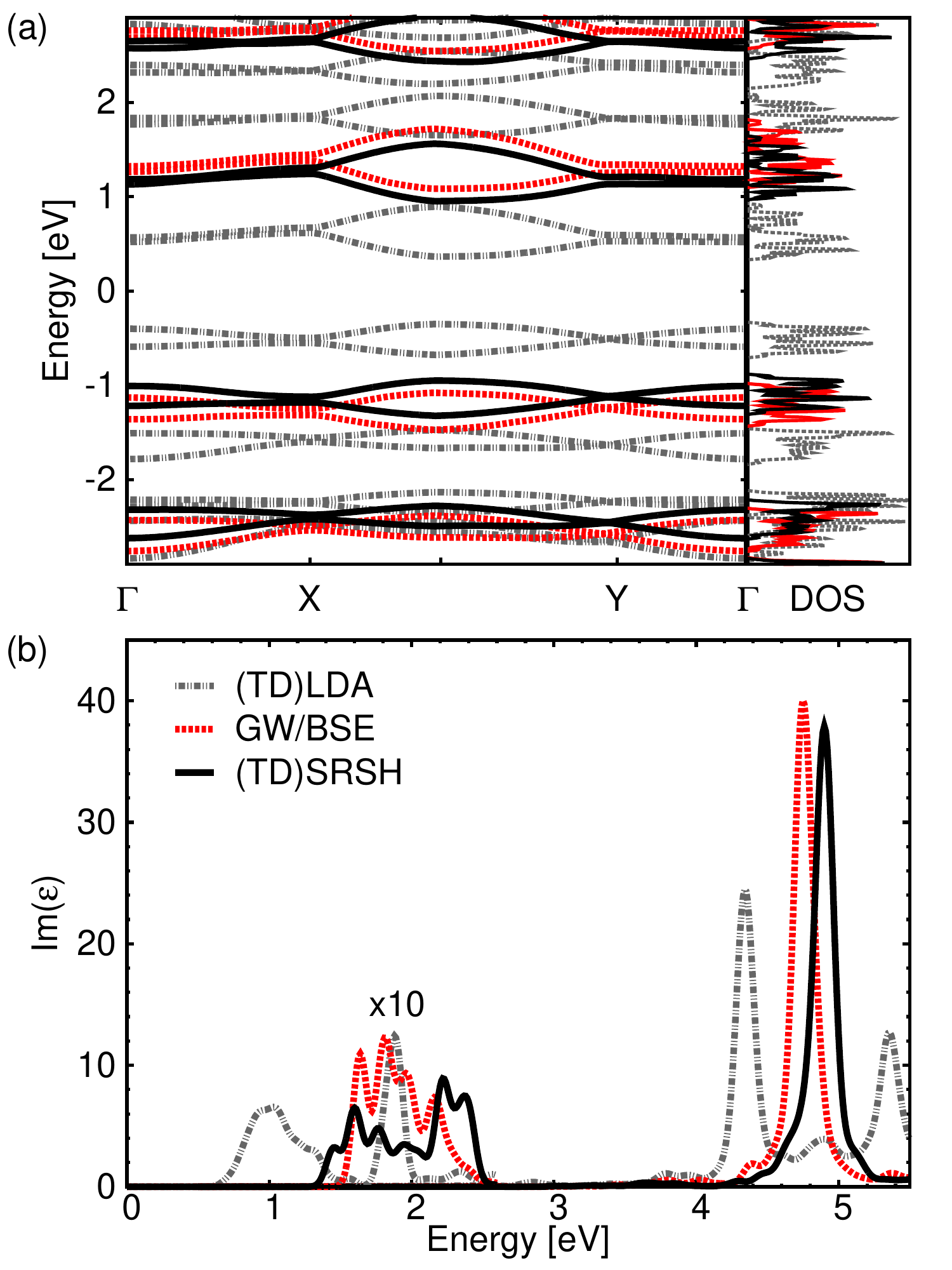}
\centering
\caption{(Color online) (a) Band-structure (left) and density of states (right) of the pentacene solid, calculated using LDA (gray, dashed lines), G$_0$W$_0$@LDA (red, dashed lines), and OT-SRSH (black, solid lines). For all methods, the middle of the bandgap is shifted to zero. (b) The imaginary part of the dielectric function of the pantacene solid, with incident light polarization averaged over the $a$, $b$, and $c$ main unit-cell axes, calculated using TDLDA (gray, dashed lines), G$_0$W$_0$/BSE (red, dashed lines), and TD-OT-SRSH (black, solid lines).  For visualization purposes, the leading absorption feature (between 0.5 to 2.5 eV) was multiplied by a factor of 10 with all computational methods used. The OT-SRSH and TD-OT-SRSH results were
obtained using the parameters $\gamma=0.16$ Bohr$^{-1}$, $\alpha=0.2$, and $\varepsilon=3.6$.  For computational details and convergence information, see the SI.}
\label{fig_pen}
\end{center}
\end{figure}

Ideally, we seek a DFT-based method where accurate one- and two-particle excitations can be read directly off of the eigenvalues of the time-independent (G)KS and the linear-response time-dependent (G)KS equations, respectively, using a single exchange-correlation functional, from which a consistent exchange-correlation potential and kernel are derived. This challenge is not met by any of the above-surveyed methods. Recently, it was met for gas-phase systems, using the optimally-tuned range separated hybrid functional (OT-RSH) approach \cite{Baer2010,Stein2010,Kronik2012}, where the long- and short- range fraction of Fock-exchange is tuned non-empirically so as to obey rigorous physical constraints. This approach, elaborated further below, was shown to yield excellent fundamental and optical gaps for molecules \cite{Refaely-Abramson2011, RSHpapers}. More recently, it has been generalized so as to provide accurate single-particle excitations for both molecules \cite{Refaely-Abramson2012,Egger2014,Nguyen2015} and molecular crystals \cite{Refaely-Abramson2013, Lueftner2014}, and it was shown to capture gap renormalization in molecular solids \cite{Refaely-Abramson2013}. Can this approach, then, resolve the long-standing challenge of providing an accurate one- and two-particle excitation spectrum in solid-state systems fully within the framework of (TD)DFT?

In this Letter, we present a solid-state OT-RSH approach that achieves just that. \red{It does so with an exchange-correlation potential and kernel that are fully consistent with the choice of the exchange-correlation energy, being its first and second functional derivative with respect to the density}. We \red{prove the accuracy of our approach by performing} non-empirical calculations for pentacene, a prototypical molecular crystal \red{and showing that it} provides excellent agreement with GW-BSE calculations. Furthermore, with one empirical parameter - set to reproduce the known fundamental gap - we again achieve results that are comparable with both GW-BSE and experiment for bulk silicon and LiF.  The approach therefore emerges as promising for \red{obtaining} photoelectron and optical properties accurately and efficiently for a broad range of extended systems.

In the range-separated hybrid approach, the Coulomb interaction is range-partitioned \cite{Leininger1997}, e.g., via: \cite{Yanai2004} 
\begin{equation}	
\frac{1}{r}=\frac{\alpha+\beta\mathrm{erf}(\gamma r)}{r}+\frac{1-[\alpha+\beta\mathrm{erf}(\gamma r)]}{r}, \label{alphabetagamma}
\end{equation}
where $\alpha$, $\beta$ and $\gamma$ are parameters, and $r$ is the inter-electron coordinate. The exchange expression corresponding to the first term on the right-hand side of Eq.\ (\ref{alphabetagamma}) is then treated as in Hartree-Fock theory; the exchange expression corresponding to the second term is treated within the Kohn-Sham framework, typically using the LDA or the generalized gradient approximation (GGA). In Eq.\ (\ref{alphabetagamma}), $\gamma$ is the range-separation parameter, i.e., it controls the range at which each of the terms dominates; $\alpha$ and $\beta$ dictate the limiting behavior of the Fock-like exchange, which tends to $\alpha/r$ for $r \rightarrow 0$ and to $(\alpha + \beta )/r$ for $r \rightarrow \infty$. The resulting exchange-correlation energy is of the form:
\begin{equation}
\begin{split}	
E_{xc}^{RSH}=(1-\alpha)E_{\mathrm{KS}x}^{SR}+\alpha E_{xx}^{SR}+
(1-(\alpha+\beta))E_{\mathrm{KS}x}^{LR}+\\
(\alpha+\beta)E_{xx}^{LR}+E_{\mathrm{KS}c},
\label{SRSH}
\end{split}
\end{equation}
where $\mathrm{KS}x$ and $\mathrm{KS}c$ denote (semi-)local KS exchange and correlation respectively, and $xx$ is a Fock-like exchange. SR and LR label short- and long-range terms, in which the Coulomb interaction is scaled using error functions. Within the GKS framework, the potential corresponding to the (semi-)local energy components is then obtained as a functional derivative, whereas the potential corresponding to the $xx$ energy components is obtained as a non-local Fock-like operator.

Here we generalize this approach to the time-dependent case, within the usual linear-response formalism of Casida \cite{Casida1995,Tretiak2003} and the Tamm-Dancoff approximation. This is achieved by coupling GKS electron-hole pairs via an exchange-correlation kernel. This leads to an eigenvalue equation in which the eigenvalues are related to optical excitation energies and the eigenvectors can be used to compute oscillator strengths, so that the complete optical absorption spectrum can be computed. The part of the TDDFT kernel originating from the Hartree and (semi-)local Kohn-Sham potential in the ground-state DFT can be expressed as:
\begin{equation}
\bra{ai}\left[\frac{1}{|r-r'|}+(1-\alpha)f_{xc}^{SR}+(1-\alpha-\beta)f_{xc}^{LR}\right]\ket{bj} \label{ker-l}
\end{equation}
where $f_{xc}^{SR}=\frac{\delta V_{xc}^{SR}}{\delta n(r)}\delta(r-r')$, $f_{xc}^{LR}=\frac{\delta V_{xc}^{LR}}{\delta n(r)}\delta(r-r')$, and $V_{xc}^{SR},V_{xc}^{LR}$ are the short- and long-range contributions of the (semi-)local KS exchange-correlation potential; and where $a,b$ and $i,j$ denote occupied and unoccupied states, respectively. The non-local exchange potential in the ground-state GKS leads to an additional term in the kernel, of the form 
\begin{equation}
-\bra{ab}\left[ \alpha\frac{\mathrm{erfc(\gamma(|r-r'|)}}{|r-r'|}+(\alpha+\beta)\frac{\mathrm{erf(\gamma(|r-r'|)}}{|r-r'|} \right] \ket{ij} \label{ker-nl}
\end{equation}
(see supplementary information (SI) for additional formal details).

As discussed above, optimal-tuning of the RSH parameters was shown to be crucial for achieving accurate description of molecular single-particle and optical excitations. This tuning procedure is challenging in the solid-state, as it involves a calculation of the system's ionization potential and electron affinity from total energy differences, a problematic procedure for periodic systems (see \cite{Vlcek2015}, and references therein). For the special case of molecular crystals, however, it was shown \cite{Refaely-Abramson2013} that predictive bandstructures can be achieved if $\gamma$ and $\alpha$ are optimally-tuned so as to obey the ionization potential theorem for an isolated molecule, with $\beta$ chosen such that $\alpha + \beta = 1/\varepsilon_0$, where $\varepsilon_0$ is the scalar dielectric constant, itself computed from first principles. We refer to this procedure as the optimally-tuned {\it screened}-RSH (OT-SRSH) approach.

To test the efficacy of this approach for optical properties, we examine pentacene, a molecular semiconducting crystal of extreme interest in organic electronics and photovoltaics. For pentacene, the optimal tuning parameters are found to be $\alpha=0.2$, $\beta=0.08$ (corresponding to $\varepsilon_0=3.6$), and $\gamma=0.16$ Bohr$^{-1}$ \cite{Refaely-Abramson2013}. With these parameters we construct the appropriate time-independent (Eq. (\ref{SRSH})) and time-dependent (Eqs.~(\ref{ker-l}),(\ref{ker-nl})) equations. These are then solved with the PARATEC \cite{Ihm1979} and BerkeleyGW \cite{Deslippe2012} codes, which we modified to handle range-separated hybrids (see SI for a complete discussion).

\begin{figure*}
\includegraphics[scale=0.55]{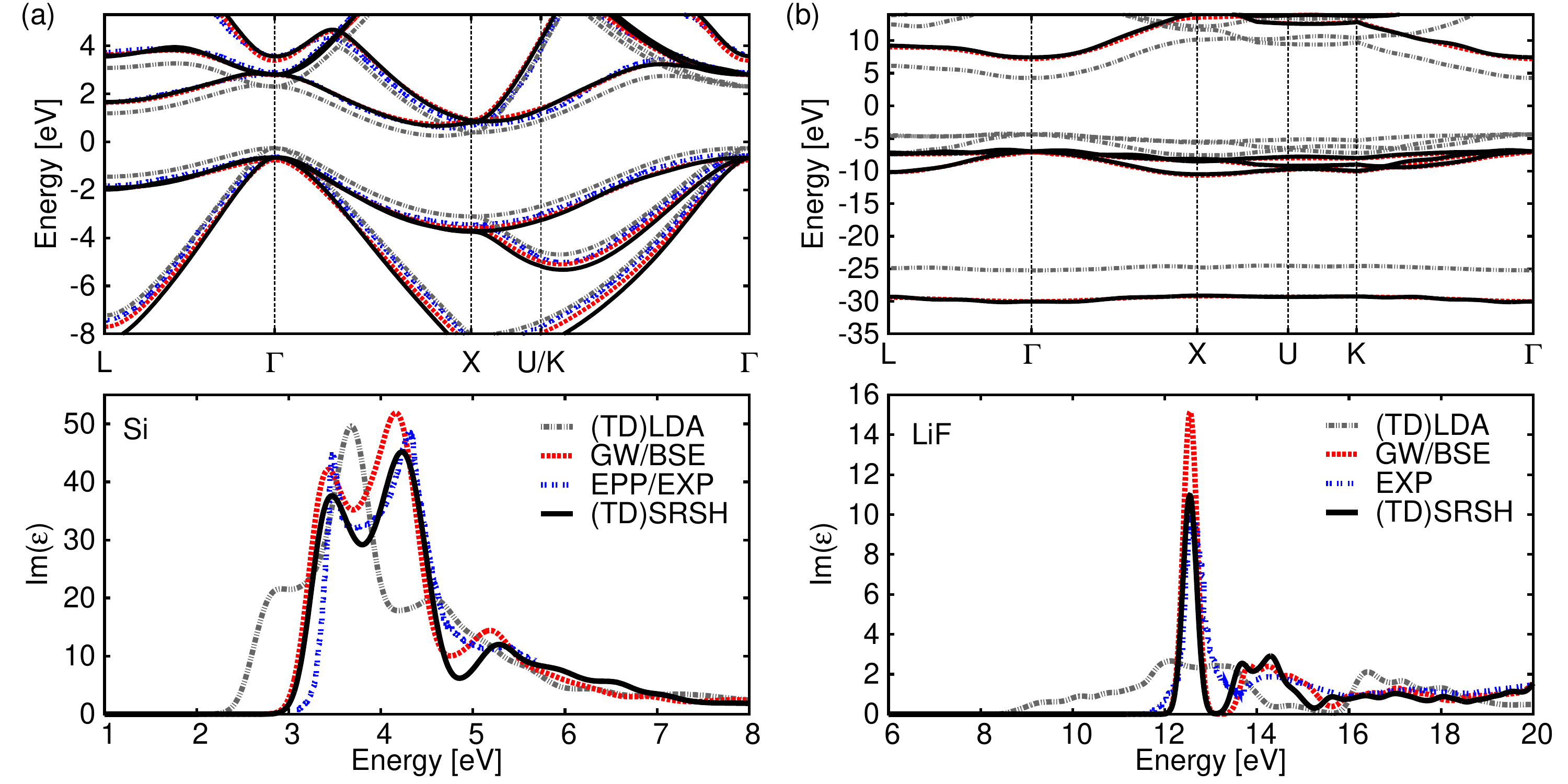}
\caption{(Color online) Top: The band-structure of bulk (a) silicon and (b) LiF, calculated using LDA (gray, dashed lines), OT-SRSH (black, solid lines), and G$_0$W$_0$@LDA (red, dashed lines). The Si bandstructure is also compared to empirical pseudopotential values taken from Ref.\ \cite{CC1976} (blue, dashed lines). For all methods, the middle of the bandgap is shifted to zero. Bottom: The imaginary part of $\varepsilon$ for (a) silicon and (b) LiF, calculated using TDLDA (gray, dashed line), TDSRSH (black, solid line) and G$_0$W$_0$/BSE (red, dashed line), and compared to experiment (blue, dashed line, Ref.\ \cite{Lautenschlager1987} for Si, and Ref.\
\cite{Roessler1967} for LiF). The optimized SRSH and TDSRSH parameters are $\gamma=0.11$ Bohr$^{-1}$, $\alpha=0.2$ , and $\varepsilon=12$ for silicon, and $\gamma=0.58$ Bohr$^{-1}$, $\alpha=0.2$, and $ \varepsilon_0=1.9$ for LiF. For computational details, see SI.}
\label{fig_si}
\end{figure*}

Figure \ref{fig_pen}~(a) shows the resulting band structure and density of states (DOS) calculated from LDA, OT-SRSH, and GW eigenvalues. The band gap predicted by OT-SRSH is 1.9~eV, in good agreement with the GW gap of 2.1~eV. As expected, both improve drastically on the LDA gap, which is 0.7 eV. Save for the slight difference in gap values, the OT-SRSH and GW band-structure and DOS are remarkably similar, demonstrating for this system that the eigenvalues of the OT-SRSH method are quantitatively useful approximations to single-quasiparticle excitation energies, as they are close to GW quasiparticle energies for several eV away from the band edges. Figure \ref{fig_pen}~(b) presents the resulting imaginary part of the dielectric function, Im($\varepsilon$), calculated using TDLDA, TD-OT-SRSH and BSE, with the response averaged over incident light polarization along the $a$, $b$ and $c$ directions of the pentacene lattice (see SI for additional details). Importantly, our TD-OT-SRSH results do as well as GW-BSE in providing the expected picture of excitonic binding, i.e., comparison of the results with and without electron-hole interactions suggests an exciton binding energy of 0.45 eV from both methods. Furthermore, the BSE result agrees well with previously reported ones, e.g. those of Refs.\ \cite{Tiago2003, Cudazzo2012, Sharifzadeh2013, *Sharifzadeh2012-2}. The first singlet excitation is predicted to be at 1.46 eV in TD-OT-SRSH and 1.64 eV in BSE. This small quantitative difference is essentially within the expected accuracy of either calculation. Moreover, this difference is consistent with that computed at the single-particle excitation level, and likely is inherited from it. In a similar manner, all presented optical excitation energies resulting from TD-OT-SRSH and BSE are very similar, with remaining differences within the desired accuracy. However, there are differences in the oscillator strength in part of the spectrum, notably at the first peak between 1.5 eV and 2.5 eV. These differences are primarily associated with the $a$-axis direction of incident light polarization. However, the scale of these differences is small when compared to the entire spectrum. (Note that this spectral range is enhanced by a factor of 10 in Fig.\ref{fig_pen}(b) to be visible.) Overall, then, the TD-OT-SRSH spectrum, while not identical to that of BSE, is in very good agreement with it.

We now turn to two prototypical covalent and ionic bulk solids: Si and LiF, respectively. For these systems, we cannot use the original tuning procedure, because it was designed for systems where inter-molecular hybridization is small. However, it still possible to set $\alpha=0.2$ as a universally useful amount of short-range exact exchange \cite{Rohrdanz2009, Egger2014} and demand $\alpha+\beta=1/\varepsilon_0$, as before. We are then left with only one parameter, the range-separation parameter, $\gamma$. Here, we simply choose $\gamma$ so as to obtain the fundamental gap. \red{We believe that future work may teach us how to obtain this parameter from first principles as well}.

As shown in the top panel of Fig \ref{fig_si}(a), the SRSH Si bandstructure is fully comparable to the canonical empirical pseudopotential work of Chelikowsky and Cohen \cite{CC1976}, as well as to GW results, whereas with LDA the gap, as expected, is too small.  
From the bottom panel we see that, in agreement with previous work \cite{Reining2002,Botti2004}, the TDLDA spectrum does not reproduce experiment satisfactorily. However, once again the TDSRSH result does almost as well as GW-BSE in predicting the experimental results, without any scissor shift or other correction operators at the single-quasiparticle level. Qualitatively, the TDSRSH lineshape is indistinguishable from that of GW-BSE. Quantitatively, the peak positions are similar to within 0.05 eV.
A similar picture emerges for LiF, as both single-particle and two-particle excitations are very close to those of GW/BSE. Specifically, both TDSRSH and BSE predict the first and largest excitation at 12.6 eV, very close to the experimental value at 12.75 eV, and the overall spectral shape is satisfying. 
Here too, then, SRSH and TDSRSH are shown to capture excitonic effects, even though the Si excitons are much more delocalized and weakly bound compared to the case of pentacene, and the LiF excitons are strongly-bound and known to be highly challenging for many of the TDDFT methods discussed in the introduction \cite{Rigamonti2015, Sharma2014,Yang2015,Trevisanutto2013}.

What are the physical origins of this success of the TDSRSH approach in the solid state? First, the ground-state calculation, being a {\it generalized} Kohn-Sham one, is capable in principle of describing quasi-particle excitations owing to the non-local potential operator. Range-separation combined with long-range dielectric screening allows us to fulfill asymptotic  potential constraints while retaining the crucial balance of short-range exchange and correlation components, thereby making such prediction sufficiently accurate also in practice. Second, the asymptotic form of the TDSRSH kernel generates the desired non-local $1/q^2$ contribution by construction. Furthermore, it is already scaled correctly via the non-empirical $1/\varepsilon_0$ parameter. 
Third, the non-locality of the TDSRSH kernel also alleviates the need for frequency-dependence, which would be necessary for bound exciton prediction with (semi-)local exchange-correlation kernels \cite{Sharma2014}. For these reasons, highly accurate single-quasiparticle and two-quasiparticle properties of extended systems can be obtained in a predictive manner within the unifying framework of DFT at computationally-modest cost. Last but not least, Onida {\it et al.} \cite{Onida2002} have already noted some time ago that ``both the Green`s functions and the TDDFT approaches profit from mutual insight.'' Here, we believe we achieve an important milestone towards that vision: on the one hand, the above work can be justified entirely from (TD)DFT reasoning. On the other hand, it is clear that the work has been motivated by the need to achieve the elegant quasi-particle picture obtained so naturally within many-body perturbation theory and that by achieving this goal we have created an effective simplified framework mimicking this picture.

In conclusion, we have presented a new approach for quantitative determination of single- and two-particle excitations in solids, based on range-separated hybrid density functional theory. The approach \red{is fully couched within the formal framework of generalized Kohn-Sham theory. Furthermore, it is based on a fully self-consistent choice of exchange-correlation energy, potential, and kernel. We have shown that, \red{with one empirical parameter at most}, it produces results} that are in quantitative agreement with those of many-body perturbation theory in the GW-BSE approximation, 
for three prototypical systems: Si, a covalent solid, LiF, an ionic solid, and pentacene, a molecular solid. In particular, it fully captures excitonic interactions for both strongly and weakly bound excitons. \red{In doing that, it answers a long-standing challenge of (TD)DFT - providing an accurate one- and two-particle excitation spectrum for solid-state systems within a fully consistent framework.}
We envision that it could emerge as a useful low-cost substitute to many-body perturbation theory, as well as provoke further developments within density functional theory. 

This work was supported by the European Research Council, the Israel Science Foundation, the United States-Israel Binational Science Foundation, the
Helmsley Foundation, and the Wolfson Foundation. S.R.A. is supported by an Adams fellowship
of the Israel Academy of Sciences and Humanities. S.S. was partially supported by the Scientific Discovery through Advanced Computing (SciDAC) Partnership program funded by US Department of Energy, Office of Science, Advanced Scientific Computing Research and Basic Energy Sciences. J.B.N. was supported by the US Department of Energy, Office
of Basic Energy Sciences, Division of Materials Sciences and Engineering (Theory FWP) under Contract No. DE-AC02-
05CH11231.  The work performed at the Molecular Foundry was also supported by the Office of Science, Office of Basic Energy Sciences, of the US Department of Energy. We thank the National Energy Research Scientific Computing center for computational resources.

\bibliographystyle{apsrev4-1}
\bibliography{tdrsh}

\begin{thebibliography}{68}%
\makeatletter
\providecommand \@ifxundefined [1]{%
 \@ifx{#1\undefined}
}%
\providecommand \@ifnum [1]{%
 \ifnum #1\expandafter \@firstoftwo
 \else \expandafter \@secondoftwo
 \fi
}%
\providecommand \@ifx [1]{%
 \ifx #1\expandafter \@firstoftwo
 \else \expandafter \@secondoftwo
 \fi
}%
\providecommand \natexlab [1]{#1}%
\providecommand \enquote  [1]{``#1''}%
\providecommand \bibnamefont  [1]{#1}%
\providecommand \bibfnamefont [1]{#1}%
\providecommand \citenamefont [1]{#1}%
\providecommand \href@noop [0]{\@secondoftwo}%
\providecommand \href [0]{\begingroup \@sanitize@url \@href}%
\providecommand \@href[1]{\@@startlink{#1}\@@href}%
\providecommand \@@href[1]{\endgroup#1\@@endlink}%
\providecommand \@sanitize@url [0]{\catcode `\\12\catcode `\$12\catcode
  `\&12\catcode `\#12\catcode `\^12\catcode `\_12\catcode `\%12\relax}%
\providecommand \@@startlink[1]{}%
\providecommand \@@endlink[0]{}%
\providecommand \url  [0]{\begingroup\@sanitize@url \@url }%
\providecommand \@url [1]{\endgroup\@href {#1}{\urlprefix }}%
\providecommand \urlprefix  [0]{URL }%
\providecommand \Eprint [0]{\href }%
\providecommand \doibase [0]{http://dx.doi.org/}%
\providecommand \selectlanguage [0]{\@gobble}%
\providecommand \bibinfo  [0]{\@secondoftwo}%
\providecommand \bibfield  [0]{\@secondoftwo}%
\providecommand \translation [1]{[#1]}%
\providecommand \BibitemOpen [0]{}%
\providecommand \bibitemStop [0]{}%
\providecommand \bibitemNoStop [0]{.\EOS\space}%
\providecommand \EOS [0]{\spacefactor3000\relax}%
\providecommand \BibitemShut  [1]{\csname bibitem#1\endcsname}%
\let\auto@bib@innerbib\@empty
\bibitem [{Sav()}]{Savoie2014}%
  \BibitemOpen
  \href@noop {} {\ }\bibinfo {note} {See, e.g., B. M. Savoie, N. E. Jackson, L.
  X. Chen, T. J. Marks, and M. A. Ratner, Acc. Chem. Res. 47, 3385
  (2014)}\BibitemShut {NoStop}%
\bibitem [{\citenamefont {Hedin}(1965)}]{Hedin1965}%
  \BibitemOpen
  \bibfield  {author} {\bibinfo {author} {\bibfnamefont {L.}~\bibnamefont
  {Hedin}},\ }\href@noop {} {\bibfield  {journal} {\bibinfo  {journal} {Phys.
  Rev.}\ }\textbf {\bibinfo {volume} {139}},\ \bibinfo {pages} {A796} (\bibinfo
  {year} {1965})}\BibitemShut {NoStop}%
\bibitem [{\citenamefont {Hybertsen}\ and\ \citenamefont
  {Louie}(1986)}]{Hybertsen1986}%
  \BibitemOpen
  \bibfield  {author} {\bibinfo {author} {\bibfnamefont {M.~S.}\ \bibnamefont
  {Hybertsen}}\ and\ \bibinfo {author} {\bibfnamefont {S.~G.}\ \bibnamefont
  {Louie}},\ }\href@noop {} {\bibfield  {journal} {\bibinfo  {journal} {Phys.
  Rev. B}\ }\textbf {\bibinfo {volume} {34}},\ \bibinfo {pages} {5390}
  (\bibinfo {year} {1986})}\BibitemShut {NoStop}%
\bibitem [{\citenamefont {Rohlfing}\ and\ \citenamefont
  {Louie}(1998)}]{Rohlfing1998}%
  \BibitemOpen
  \bibfield  {author} {\bibinfo {author} {\bibfnamefont {M.}~\bibnamefont
  {Rohlfing}}\ and\ \bibinfo {author} {\bibfnamefont {S.~G.}\ \bibnamefont
  {Louie}},\ }\href@noop {} {\bibfield  {journal} {\bibinfo  {journal} {Phys.
  Rev. Lett.}\ }\textbf {\bibinfo {volume} {81}},\ \bibinfo {pages} {2312}
  (\bibinfo {year} {1998})}\BibitemShut {NoStop}%
\bibitem [{\citenamefont {Rohlfing}\ and\ \citenamefont
  {Louie}(2000)}]{Rohlfing2000}%
  \BibitemOpen
  \bibfield  {author} {\bibinfo {author} {\bibfnamefont {M.}~\bibnamefont
  {Rohlfing}}\ and\ \bibinfo {author} {\bibfnamefont {S.~G.}\ \bibnamefont
  {Louie}},\ }\href@noop {} {\bibfield  {journal} {\bibinfo  {journal} {Phys.
  Rev. B}\ }\textbf {\bibinfo {volume} {62}},\ \bibinfo {pages} {4927}
  (\bibinfo {year} {2000})}\BibitemShut {NoStop}%
\bibitem [{\citenamefont {Strinati}(1982)}]{Strinati1982}%
  \BibitemOpen
  \bibfield  {author} {\bibinfo {author} {\bibfnamefont {G.}~\bibnamefont
  {Strinati}},\ }\href@noop {} {\bibfield  {journal} {\bibinfo  {journal}
  {Phys. Rev. Lett.}\ }\textbf {\bibinfo {volume} {49}},\ \bibinfo {pages}
  {1519} (\bibinfo {year} {1982})}\BibitemShut {NoStop}%
\bibitem [{\citenamefont {Strinati}(1984)}]{Strinati1984}%
  \BibitemOpen
  \bibfield  {author} {\bibinfo {author} {\bibfnamefont {G.}~\bibnamefont
  {Strinati}},\ }\href@noop {} {\bibfield  {journal} {\bibinfo  {journal}
  {Phys. Rev. B}\ }\textbf {\bibinfo {volume} {29}},\ \bibinfo {pages} {5718}
  (\bibinfo {year} {1984})}\BibitemShut {NoStop}%
\bibitem [{\citenamefont {Gross}\ and\ \citenamefont
  {Dreizler}(1995)}]{DreizlerGross}%
  \BibitemOpen
  \bibfield  {author} {\bibinfo {author} {\bibfnamefont {E.~K.~U.}\
  \bibnamefont {Gross}}\ and\ \bibinfo {author} {\bibfnamefont {R.~M.}\
  \bibnamefont {Dreizler}},\ }\href@noop {} {\emph {\bibinfo {title} {Density
  Functional Theory}}}\ (\bibinfo  {publisher} {Plenum Press},\ \bibinfo
  {address} {New York},\ \bibinfo {year} {1995})\BibitemShut {NoStop}%
\bibitem [{\citenamefont {Parr}\ and\ \citenamefont {Yang}(1989)}]{ParrYang}%
  \BibitemOpen
  \bibfield  {author} {\bibinfo {author} {\bibfnamefont {R.~G.}\ \bibnamefont
  {Parr}}\ and\ \bibinfo {author} {\bibfnamefont {W.}~\bibnamefont {Yang}},\
  }\href@noop {} {\emph {\bibinfo {title} {Density Functional Theory of Atoms
  and Molecules}}}\ (\bibinfo  {publisher} {Oxford University Press},\ \bibinfo
  {address} {New York},\ \bibinfo {year} {1989})\BibitemShut {NoStop}%
\bibitem [{\citenamefont {Marques}\ \emph {et~al.}(2012)\citenamefont
  {Marques}, \citenamefont {Maitra}, \citenamefont {Nogueria}, \citenamefont
  {Gross},\ and\ \citenamefont {Rubiu}}]{Marques2012}%
  \BibitemOpen
  \bibinfo {editor} {\bibfnamefont {M.~A.~L.}\ \bibnamefont {Marques}},
  \bibinfo {editor} {\bibfnamefont {N.}~\bibnamefont {Maitra}}, \bibinfo
  {editor} {\bibfnamefont {F.}~\bibnamefont {Nogueria}}, \bibinfo {editor}
  {\bibfnamefont {E.~K.~U.}\ \bibnamefont {Gross}}, \ and\ \bibinfo {editor}
  {\bibfnamefont {A.}~\bibnamefont {Rubiu}},\ eds.,\ \href@noop {} {\emph
  {\bibinfo {title} {Fundamentals of Time-Dependent Functional Theory}}}\
  (\bibinfo  {publisher} {Springer-Verlag},\ \bibinfo {address} {Berlin},\
  \bibinfo {year} {2012})\BibitemShut {NoStop}%
\bibitem [{\citenamefont {Casida}(1995)}]{Casida1995}%
  \BibitemOpen
  \bibfield  {author} {\bibinfo {author} {\bibfnamefont {M.~E.}\ \bibnamefont
  {Casida}},\ }in\ \href@noop {} {\emph {\bibinfo {booktitle} {Recent Advances
  in Density-Functional Methods part I}}},\ \bibinfo {editor} {edited by\
  \bibinfo {editor} {\bibfnamefont {D.~P.}\ \bibnamefont {Chong}}}\ (\bibinfo
  {publisher} {World Scientific},\ \bibinfo {address} {Singapore},\ \bibinfo
  {year} {1995})\ p.\ \bibinfo {pages} {155}\BibitemShut {NoStop}%
\bibitem [{\citenamefont {Burke}\ \emph {et~al.}(2005)\citenamefont {Burke},
  \citenamefont {Werschnik},\ and\ \citenamefont {Gross}}]{Burke2005}%
  \BibitemOpen
  \bibfield  {author} {\bibinfo {author} {\bibfnamefont {K.}~\bibnamefont
  {Burke}}, \bibinfo {author} {\bibfnamefont {J.}~\bibnamefont {Werschnik}}, \
  and\ \bibinfo {author} {\bibfnamefont {E.~K.~U.}\ \bibnamefont {Gross}},\
  }\href@noop {} {\bibfield  {journal} {\bibinfo  {journal} {J. Chem. Phys.}\
  }\textbf {\bibinfo {volume} {123}},\ \bibinfo {pages} {062206} (\bibinfo
  {year} {2005})}\BibitemShut {NoStop}%
\bibitem [{\citenamefont {Ullrich}(2012)}]{Ullrich_book}%
  \BibitemOpen
  \bibfield  {author} {\bibinfo {author} {\bibfnamefont {C.~A.}\ \bibnamefont
  {Ullrich}},\ }\href@noop {} {\emph {\bibinfo {title} {Time-Dependent
  Density-Functional Theory - Concepts and Applications}}}\ (\bibinfo
  {publisher} {Oxford University Press},\ \bibinfo {address} {New York},\
  \bibinfo {year} {2012})\BibitemShut {NoStop}%
\bibitem [{\citenamefont {Onida}\ \emph {et~al.}(2002)\citenamefont {Onida},
  \citenamefont {Reining},\ and\ \citenamefont {Rubio}}]{Onida2002}%
  \BibitemOpen
  \bibfield  {author} {\bibinfo {author} {\bibfnamefont {G.}~\bibnamefont
  {Onida}}, \bibinfo {author} {\bibfnamefont {L.}~\bibnamefont {Reining}}, \
  and\ \bibinfo {author} {\bibfnamefont {A.}~\bibnamefont {Rubio}},\
  }\href@noop {} {\bibfield  {journal} {\bibinfo  {journal} {Rev. Mod. Phys.}\
  }\textbf {\bibinfo {volume} {74}},\ \bibinfo {pages} {601} (\bibinfo {year}
  {2002})}\BibitemShut {NoStop}%
\bibitem [{\citenamefont {K$\ddot{\mbox{u}}$mmel}\ and\ \citenamefont
  {Kronik}(2008)}]{Kummel2008}%
  \BibitemOpen
  \bibfield  {author} {\bibinfo {author} {\bibfnamefont {S.}~\bibnamefont
  {K$\ddot{\mbox{u}}$mmel}}\ and\ \bibinfo {author} {\bibfnamefont
  {L.}~\bibnamefont {Kronik}},\ }\href@noop {} {\bibfield  {journal} {\bibinfo
  {journal} {Rev. Mod. Phys.}\ }\textbf {\bibinfo {volume} {80}},\ \bibinfo
  {pages} {3} (\bibinfo {year} {2008})}\BibitemShut {NoStop}%
\bibitem [{\citenamefont {Kronik}\ \emph {et~al.}(2012)\citenamefont {Kronik},
  \citenamefont {Stein}, \citenamefont {Refaely-Abramson},\ and\ \citenamefont
  {Baer}}]{Kronik2012}%
  \BibitemOpen
  \bibfield  {author} {\bibinfo {author} {\bibfnamefont {L.}~\bibnamefont
  {Kronik}}, \bibinfo {author} {\bibfnamefont {T.}~\bibnamefont {Stein}},
  \bibinfo {author} {\bibfnamefont {S.}~\bibnamefont {Refaely-Abramson}}, \
  and\ \bibinfo {author} {\bibfnamefont {R.}~\bibnamefont {Baer}},\ }\href@noop
  {} {\bibfield  {journal} {\bibinfo  {journal} {J. Chem. Theory Comput.}\
  }\textbf {\bibinfo {volume} {8}},\ \bibinfo {pages} {1515} (\bibinfo {year}
  {2012})}\BibitemShut {NoStop}%
\bibitem [{\citenamefont {Salzner}\ \emph {et~al.}(1997)\citenamefont
  {Salzner}, \citenamefont {Lagowski}, \citenamefont {Pickup},\ and\
  \citenamefont {Piorier}}]{Salzner1997}%
  \BibitemOpen
  \bibfield  {author} {\bibinfo {author} {\bibfnamefont {U.}~\bibnamefont
  {Salzner}}, \bibinfo {author} {\bibfnamefont {J.~B.}\ \bibnamefont
  {Lagowski}}, \bibinfo {author} {\bibfnamefont {P.~G.}\ \bibnamefont
  {Pickup}}, \ and\ \bibinfo {author} {\bibfnamefont {R.~A.}\ \bibnamefont
  {Piorier}},\ }\href@noop {} {\bibfield  {journal} {\bibinfo  {journal} {J.
  Comput. Chem.}\ }\textbf {\bibinfo {volume} {18}},\ \bibinfo {pages} {1943}
  (\bibinfo {year} {1997})}\BibitemShut {NoStop}%
\bibitem [{\citenamefont {Chong}\ \emph {et~al.}(2002)\citenamefont {Chong},
  \citenamefont {Gritsenko},\ and\ \citenamefont {Baerends}}]{Chong2002}%
  \BibitemOpen
  \bibfield  {author} {\bibinfo {author} {\bibfnamefont {D.~P.}\ \bibnamefont
  {Chong}}, \bibinfo {author} {\bibfnamefont {O.~V.}\ \bibnamefont
  {Gritsenko}}, \ and\ \bibinfo {author} {\bibfnamefont {E.~J.}\ \bibnamefont
  {Baerends}},\ }\href@noop {} {\bibfield  {journal} {\bibinfo  {journal} {J.
  Chem. Phys.}\ }\textbf {\bibinfo {volume} {116}},\ \bibinfo {pages} {1760}
  (\bibinfo {year} {2002})}\BibitemShut {NoStop}%
\bibitem [{\citenamefont {Baerends}\ \emph {et~al.}(2013)\citenamefont
  {Baerends}, \citenamefont {Gritsenko},\ and\ \citenamefont {{van
  Meer}}}]{Baerends2013}%
  \BibitemOpen
  \bibfield  {author} {\bibinfo {author} {\bibfnamefont {E.~J.}\ \bibnamefont
  {Baerends}}, \bibinfo {author} {\bibfnamefont {O.~V.}\ \bibnamefont
  {Gritsenko}}, \ and\ \bibinfo {author} {\bibfnamefont {R.}~\bibnamefont {{van
  Meer}}},\ }\href@noop {} {\bibfield  {journal} {\bibinfo  {journal} {Phys.
  Chem. Chem. Phys.}\ }\textbf {\bibinfo {volume} {15}},\ \bibinfo {pages}
  {16408} (\bibinfo {year} {2013})}\BibitemShut {NoStop}%
\bibitem [{\citenamefont {Kronik}\ and\ \citenamefont
  {K\"{u}mmel}(2014)}]{Kronik2014}%
  \BibitemOpen
  \bibfield  {author} {\bibinfo {author} {\bibfnamefont {L.}~\bibnamefont
  {Kronik}}\ and\ \bibinfo {author} {\bibfnamefont {S.}~\bibnamefont
  {K\"{u}mmel}},\ }\href@noop {} {\bibfield  {journal} {\bibinfo  {journal}
  {Top. Curr. Chem}\ }\textbf {\bibinfo {volume} {347}},\ \bibinfo {pages}
  {137} (\bibinfo {year} {2014})}\BibitemShut {NoStop}%
\bibitem [{\citenamefont {Izmaylov}\ and\ \citenamefont
  {Scuseria}(2008)}]{Izmaylov2008}%
  \BibitemOpen
  \bibfield  {author} {\bibinfo {author} {\bibfnamefont {A.~F.}\ \bibnamefont
  {Izmaylov}}\ and\ \bibinfo {author} {\bibfnamefont {G.~E.}\ \bibnamefont
  {Scuseria}},\ }\href@noop {} {\bibfield  {journal} {\bibinfo  {journal} {J.
  Chem. Phys.}\ }\textbf {\bibinfo {volume} {129}},\ \bibinfo {pages} {034101}
  (\bibinfo {year} {2008})}\BibitemShut {NoStop}%
\bibitem [{\citenamefont {Sharma}\ \emph {et~al.}(2014)\citenamefont {Sharma},
  \citenamefont {Dewhurst},\ and\ \citenamefont {Gross}}]{Sharma2014}%
  \BibitemOpen
  \bibfield  {author} {\bibinfo {author} {\bibfnamefont {S.}~\bibnamefont
  {Sharma}}, \bibinfo {author} {\bibfnamefont {J.~K.}\ \bibnamefont
  {Dewhurst}}, \ and\ \bibinfo {author} {\bibfnamefont {E.~K.~U.}\ \bibnamefont
  {Gross}},\ }\href@noop {} {\bibfield  {journal} {\bibinfo  {journal} {Top.
  Curr. Chem.}\ }\textbf {\bibinfo {volume} {347}},\ \bibinfo {pages} {235}
  (\bibinfo {year} {2014})}\BibitemShut {NoStop}%
\bibitem [{\citenamefont {Ullrich}\ and\ \citenamefont {hui
  Yang}(2014)}]{Ullrich2014}%
  \BibitemOpen
  \bibfield  {author} {\bibinfo {author} {\bibfnamefont {C.~A.}\ \bibnamefont
  {Ullrich}}\ and\ \bibinfo {author} {\bibfnamefont {Z.}~\bibnamefont {hui
  Yang}},\ }\href@noop {} {\bibfield  {journal} {\bibinfo  {journal} {Braz. J.
  Phys.}\ }\textbf {\bibinfo {volume} {44}},\ \bibinfo {pages} {154} (\bibinfo
  {year} {2014})}\BibitemShut {NoStop}%
\bibitem [{\citenamefont {Gonze}\ \emph {et~al.}(1995)\citenamefont {Gonze},
  \citenamefont {Ghosez},\ and\ \citenamefont {Godby}}]{Gonze1995}%
  \BibitemOpen
  \bibfield  {author} {\bibinfo {author} {\bibfnamefont {X.}~\bibnamefont
  {Gonze}}, \bibinfo {author} {\bibfnamefont {P.}~\bibnamefont {Ghosez}}, \
  and\ \bibinfo {author} {\bibfnamefont {R.~W.}\ \bibnamefont {Godby}},\
  }\href@noop {} {\bibfield  {journal} {\bibinfo  {journal} {Phys. Rev. Lett.}\
  }\textbf {\bibinfo {volume} {74}},\ \bibinfo {pages} {4035} (\bibinfo {year}
  {1995})}\BibitemShut {NoStop}%
\bibitem [{\citenamefont {Ghosez}\ \emph {et~al.}(1997)\citenamefont {Ghosez},
  \citenamefont {Gonze},\ and\ \citenamefont {Godby}}]{Ghosez1997}%
  \BibitemOpen
  \bibfield  {author} {\bibinfo {author} {\bibfnamefont {P.}~\bibnamefont
  {Ghosez}}, \bibinfo {author} {\bibfnamefont {X.}~\bibnamefont {Gonze}}, \
  and\ \bibinfo {author} {\bibfnamefont {R.~W.}\ \bibnamefont {Godby}},\
  }\href@noop {} {\bibfield  {journal} {\bibinfo  {journal} {Phys. Rev. B}\
  }\textbf {\bibinfo {volume} {56}},\ \bibinfo {pages} {12811} (\bibinfo {year}
  {1997})}\BibitemShut {NoStop}%
\bibitem [{\citenamefont {Kim}\ and\ \citenamefont
  {G\"{o}rling}(2002)}]{Kim2002}%
  \BibitemOpen
  \bibfield  {author} {\bibinfo {author} {\bibfnamefont {Y.-H.}\ \bibnamefont
  {Kim}}\ and\ \bibinfo {author} {\bibfnamefont {A.}~\bibnamefont
  {G\"{o}rling}},\ }\href@noop {} {\bibfield  {journal} {\bibinfo  {journal}
  {Phys. Rev. Lett.}\ }\textbf {\bibinfo {volume} {89}},\ \bibinfo {pages}
  {096402} (\bibinfo {year} {2002})}\BibitemShut {NoStop}%
\bibitem [{\citenamefont {Reining}\ \emph {et~al.}(2002)\citenamefont
  {Reining}, \citenamefont {Olevano}, \citenamefont {Rubio},\ and\
  \citenamefont {Onida}}]{Reining2002}%
  \BibitemOpen
  \bibfield  {author} {\bibinfo {author} {\bibfnamefont {L.}~\bibnamefont
  {Reining}}, \bibinfo {author} {\bibfnamefont {V.}~\bibnamefont {Olevano}},
  \bibinfo {author} {\bibfnamefont {A.}~\bibnamefont {Rubio}}, \ and\ \bibinfo
  {author} {\bibfnamefont {G.}~\bibnamefont {Onida}},\ }\href@noop {}
  {\bibfield  {journal} {\bibinfo  {journal} {Phys. Rev. Lett.}\ }\textbf
  {\bibinfo {volume} {88}},\ \bibinfo {pages} {066404} (\bibinfo {year}
  {2002})}\BibitemShut {NoStop}%
\bibitem [{\citenamefont {Botti}\ \emph {et~al.}(2004)\citenamefont {Botti},
  \citenamefont {Sottile}, \citenamefont {Vast}, \citenamefont {Olevano},
  \citenamefont {Reining}, \citenamefont {Weissker}, \citenamefont {Rubio},
  \citenamefont {Onida}, \citenamefont {Del-Sole},\ and\ \citenamefont
  {Godby}}]{Botti2004}%
  \BibitemOpen
  \bibfield  {author} {\bibinfo {author} {\bibfnamefont {S.}~\bibnamefont
  {Botti}}, \bibinfo {author} {\bibfnamefont {F.}~\bibnamefont {Sottile}},
  \bibinfo {author} {\bibfnamefont {N.}~\bibnamefont {Vast}}, \bibinfo {author}
  {\bibfnamefont {V.}~\bibnamefont {Olevano}}, \bibinfo {author} {\bibfnamefont
  {L.}~\bibnamefont {Reining}}, \bibinfo {author} {\bibfnamefont {H.-C.}\
  \bibnamefont {Weissker}}, \bibinfo {author} {\bibfnamefont {A.}~\bibnamefont
  {Rubio}}, \bibinfo {author} {\bibfnamefont {G.}~\bibnamefont {Onida}},
  \bibinfo {author} {\bibfnamefont {R.}~\bibnamefont {Del-Sole}}, \ and\
  \bibinfo {author} {\bibfnamefont {R.~W.}\ \bibnamefont {Godby}},\ }\href@noop
  {} {\bibfield  {journal} {\bibinfo  {journal} {Phys. Rev. B}\ }\textbf
  {\bibinfo {volume} {69}},\ \bibinfo {pages} {155112} (\bibinfo {year}
  {2004})}\BibitemShut {NoStop}%
\bibitem [{\citenamefont {Trevisanutto}\ \emph {et~al.}(2015)\citenamefont
  {Trevisanutto}, \citenamefont {Terentjevs}, \citenamefont {Constantin},
  \citenamefont {Olevano},\ and\ \citenamefont {Sala}}]{Trevisanutto2013}%
  \BibitemOpen
  \bibfield  {author} {\bibinfo {author} {\bibfnamefont {P.~E.}\ \bibnamefont
  {Trevisanutto}}, \bibinfo {author} {\bibfnamefont {A.}~\bibnamefont
  {Terentjevs}}, \bibinfo {author} {\bibfnamefont {L.~A.}\ \bibnamefont
  {Constantin}}, \bibinfo {author} {\bibfnamefont {V.}~\bibnamefont {Olevano}},
  \ and\ \bibinfo {author} {\bibfnamefont {F.~D.}\ \bibnamefont {Sala}},\
  }\href@noop {} {\bibfield  {journal} {\bibinfo  {journal} {Phys. Rev. Lett.}\
  }\textbf {\bibinfo {volume} {114}},\ \bibinfo {pages} {146402} (\bibinfo
  {year} {2015})}\BibitemShut {NoStop}%
\bibitem [{\citenamefont {Sharma}\ \emph {et~al.}(2011)\citenamefont {Sharma},
  \citenamefont {Dewhurst}, \citenamefont {Sanna},\ and\ \citenamefont
  {Gross}}]{Sharma2011}%
  \BibitemOpen
  \bibfield  {author} {\bibinfo {author} {\bibfnamefont {S.}~\bibnamefont
  {Sharma}}, \bibinfo {author} {\bibfnamefont {J.~K.}\ \bibnamefont
  {Dewhurst}}, \bibinfo {author} {\bibfnamefont {A.}~\bibnamefont {Sanna}}, \
  and\ \bibinfo {author} {\bibfnamefont {E.~K.~U.}\ \bibnamefont {Gross}},\
  }\href@noop {} {\bibfield  {journal} {\bibinfo  {journal} {Phys. Rev. Lett.}\
  }\textbf {\bibinfo {volume} {107}},\ \bibinfo {pages} {186401} (\bibinfo
  {year} {2011})}\BibitemShut {NoStop}%
\bibitem [{\citenamefont {Rigamonti}\ \emph {et~al.}(2015)\citenamefont
  {Rigamonti}, \citenamefont {Botti}, \citenamefont {Veniard}, \citenamefont
  {Draxl}, \citenamefont {Reining},\ and\ \citenamefont
  {Sottile}}]{Rigamonti2015}%
  \BibitemOpen
  \bibfield  {author} {\bibinfo {author} {\bibfnamefont {S.}~\bibnamefont
  {Rigamonti}}, \bibinfo {author} {\bibfnamefont {S.}~\bibnamefont {Botti}},
  \bibinfo {author} {\bibfnamefont {V.}~\bibnamefont {Veniard}}, \bibinfo
  {author} {\bibfnamefont {C.}~\bibnamefont {Draxl}}, \bibinfo {author}
  {\bibfnamefont {L.}~\bibnamefont {Reining}}, \ and\ \bibinfo {author}
  {\bibfnamefont {F.}~\bibnamefont {Sottile}},\ }\href@noop {} {\bibfield
  {journal} {\bibinfo  {journal} {Phys. Rev. Lett.}\ }\textbf {\bibinfo
  {volume} {114}},\ \bibinfo {pages} {146402} (\bibinfo {year}
  {2015})}\BibitemShut {NoStop}%
\bibitem [{\citenamefont {Sottile}\ \emph {et~al.}(2003)\citenamefont
  {Sottile}, \citenamefont {Olevano},\ and\ \citenamefont
  {Reining}}]{Sottile2003}%
  \BibitemOpen
  \bibfield  {author} {\bibinfo {author} {\bibfnamefont {F.}~\bibnamefont
  {Sottile}}, \bibinfo {author} {\bibfnamefont {V.}~\bibnamefont {Olevano}}, \
  and\ \bibinfo {author} {\bibfnamefont {L.}~\bibnamefont {Reining}},\
  }\href@noop {} {\bibfield  {journal} {\bibinfo  {journal} {Phys. Rev. Lett.}\
  }\textbf {\bibinfo {volume} {91}},\ \bibinfo {pages} {056402} (\bibinfo
  {year} {2003})}\BibitemShut {NoStop}%
\bibitem [{\citenamefont {Marini}\ \emph {et~al.}(2003)\citenamefont {Marini},
  \citenamefont {Del-Sole},\ and\ \citenamefont {Rubio}}]{Marini2003}%
  \BibitemOpen
  \bibfield  {author} {\bibinfo {author} {\bibfnamefont {A.}~\bibnamefont
  {Marini}}, \bibinfo {author} {\bibfnamefont {R.}~\bibnamefont {Del-Sole}}, \
  and\ \bibinfo {author} {\bibfnamefont {A.}~\bibnamefont {Rubio}},\
  }\href@noop {} {\bibfield  {journal} {\bibinfo  {journal} {Phys. Rev. Lett.}\
  }\textbf {\bibinfo {volume} {91}},\ \bibinfo {pages} {256402} (\bibinfo
  {year} {2003})}\BibitemShut {NoStop}%
\bibitem [{\citenamefont {Adragna}\ \emph {et~al.}(2003)\citenamefont
  {Adragna}, \citenamefont {Del-Sole},\ and\ \citenamefont
  {Marini}}]{Adragna2003}%
  \BibitemOpen
  \bibfield  {author} {\bibinfo {author} {\bibfnamefont {G.}~\bibnamefont
  {Adragna}}, \bibinfo {author} {\bibfnamefont {R.}~\bibnamefont {Del-Sole}}, \
  and\ \bibinfo {author} {\bibfnamefont {A.}~\bibnamefont {Marini}},\
  }\href@noop {} {\bibfield  {journal} {\bibinfo  {journal} {Phys. Rev. B}\
  }\textbf {\bibinfo {volume} {68}},\ \bibinfo {pages} {165108} (\bibinfo
  {year} {2003})}\BibitemShut {NoStop}%
\bibitem [{\citenamefont {Seidl}\ \emph {et~al.}(1996)\citenamefont {Seidl},
  \citenamefont {G\"orling}, \citenamefont {Vogl}, \citenamefont {Majewski},\
  and\ \citenamefont {Levy}}]{Seidl1996}%
  \BibitemOpen
  \bibfield  {author} {\bibinfo {author} {\bibfnamefont {A.}~\bibnamefont
  {Seidl}}, \bibinfo {author} {\bibfnamefont {A.}~\bibnamefont {G\"orling}},
  \bibinfo {author} {\bibfnamefont {P.}~\bibnamefont {Vogl}}, \bibinfo {author}
  {\bibfnamefont {J.~A.}\ \bibnamefont {Majewski}}, \ and\ \bibinfo {author}
  {\bibfnamefont {M.}~\bibnamefont {Levy}},\ }\href@noop {} {\bibfield
  {journal} {\bibinfo  {journal} {Phys. Rev. B}\ }\textbf {\bibinfo {volume}
  {53}},\ \bibinfo {pages} {3764} (\bibinfo {year} {1996})}\BibitemShut
  {NoStop}%
\bibitem [{\citenamefont {Heyd}\ \emph {et~al.}(2003)\citenamefont {Heyd},
  \citenamefont {Scuseria},\ and\ \citenamefont {Ernzerhof}}]{Heyd2003}%
  \BibitemOpen
  \bibfield  {author} {\bibinfo {author} {\bibfnamefont {J.}~\bibnamefont
  {Heyd}}, \bibinfo {author} {\bibfnamefont {G.~E.}\ \bibnamefont {Scuseria}},
  \ and\ \bibinfo {author} {\bibfnamefont {M.~J.}\ \bibnamefont {Ernzerhof}},\
  }\href@noop {} {\bibfield  {journal} {\bibinfo  {journal} {J. Chem. Phys.}\
  }\textbf {\bibinfo {volume} {118}},\ \bibinfo {pages} {8207} (\bibinfo {year}
  {2003})}\BibitemShut {NoStop}%
\bibitem [{\citenamefont {Heyd}\ \emph {et~al.}(2006)\citenamefont {Heyd},
  \citenamefont {Scuseria},\ and\ \citenamefont {Ernzerhof}}]{Heyd2006}%
  \BibitemOpen
  \bibfield  {author} {\bibinfo {author} {\bibfnamefont {J.}~\bibnamefont
  {Heyd}}, \bibinfo {author} {\bibfnamefont {G.~E.}\ \bibnamefont {Scuseria}},
  \ and\ \bibinfo {author} {\bibfnamefont {M.~J.}\ \bibnamefont {Ernzerhof}},\
  }\href@noop {} {\bibfield  {journal} {\bibinfo  {journal} {J. Chem. Phys.}\
  }\textbf {\bibinfo {volume} {124}},\ \bibinfo {pages} {219906} (\bibinfo
  {year} {2006})}\BibitemShut {NoStop}%
\bibitem [{\citenamefont {Paier}\ \emph {et~al.}(2008)\citenamefont {Paier},
  \citenamefont {Marsman},\ and\ \citenamefont {Kresse}}]{Paier2008}%
  \BibitemOpen
  \bibfield  {author} {\bibinfo {author} {\bibfnamefont {J.}~\bibnamefont
  {Paier}}, \bibinfo {author} {\bibfnamefont {M.}~\bibnamefont {Marsman}}, \
  and\ \bibinfo {author} {\bibfnamefont {G.}~\bibnamefont {Kresse}},\
  }\href@noop {} {\bibfield  {journal} {\bibinfo  {journal} {Phys. Rev. B}\
  }\textbf {\bibinfo {volume} {78}},\ \bibinfo {pages} {121201(R)} (\bibinfo
  {year} {2008})}\BibitemShut {NoStop}%
\bibitem [{\citenamefont {Becke}(1993)}]{Becke1993}%
  \BibitemOpen
  \bibfield  {author} {\bibinfo {author} {\bibfnamefont {A.~D.}\ \bibnamefont
  {Becke}},\ }\href@noop {} {\bibfield  {journal} {\bibinfo  {journal} {J.
  Chem. Phys.}\ }\textbf {\bibinfo {volume} {98}},\ \bibinfo {pages} {5648}
  (\bibinfo {year} {1993})}\BibitemShut {NoStop}%
\bibitem [{\citenamefont {Stephens}\ \emph {et~al.}(1994)\citenamefont
  {Stephens}, \citenamefont {Devlin}, \citenamefont {Chabalowski},\ and\
  \citenamefont {Frisch}}]{Stephens1994}%
  \BibitemOpen
  \bibfield  {author} {\bibinfo {author} {\bibfnamefont {P.~J.}\ \bibnamefont
  {Stephens}}, \bibinfo {author} {\bibfnamefont {F.~J.}\ \bibnamefont
  {Devlin}}, \bibinfo {author} {\bibfnamefont {C.~F.}\ \bibnamefont
  {Chabalowski}}, \ and\ \bibinfo {author} {\bibfnamefont {M.~J.}\ \bibnamefont
  {Frisch}},\ }\href@noop {} {\bibfield  {journal} {\bibinfo  {journal} {J.
  Phys. Chem.}\ }\textbf {\bibinfo {volume} {98}},\ \bibinfo {pages} {11623}
  (\bibinfo {year} {1994})}\BibitemShut {NoStop}%
\bibitem [{\citenamefont {Bernasconi}\ \emph {et~al.}(2011)\citenamefont
  {Bernasconi}, \citenamefont {Tomi\'{c}}, \citenamefont {Ferrero},
  \citenamefont {R\'{e}rat}, \citenamefont {Orlando}, \citenamefont {Dovesi},\
  and\ \citenamefont {Harrison}}]{Bernasconi2011}%
  \BibitemOpen
  \bibfield  {author} {\bibinfo {author} {\bibfnamefont {L.}~\bibnamefont
  {Bernasconi}}, \bibinfo {author} {\bibfnamefont {S.}~\bibnamefont
  {Tomi\'{c}}}, \bibinfo {author} {\bibfnamefont {M.}~\bibnamefont {Ferrero}},
  \bibinfo {author} {\bibfnamefont {M.}~\bibnamefont {R\'{e}rat}}, \bibinfo
  {author} {\bibfnamefont {R.}~\bibnamefont {Orlando}}, \bibinfo {author}
  {\bibfnamefont {R.}~\bibnamefont {Dovesi}}, \ and\ \bibinfo {author}
  {\bibfnamefont {N.~M.}\ \bibnamefont {Harrison}},\ }\href@noop {} {\bibfield
  {journal} {\bibinfo  {journal} {Phys. Rev. B}\ }\textbf {\bibinfo {volume}
  {83}},\ \bibinfo {pages} {195325} (\bibinfo {year} {2011})}\BibitemShut
  {NoStop}%
\bibitem [{\citenamefont {Tomi\'{c}}\ \emph {et~al.}(2014)\citenamefont
  {Tomi\'{c}}, \citenamefont {Bernasconi}, \citenamefont {Searle},\ and\
  \citenamefont {Harrison}}]{Tomic2014}%
  \BibitemOpen
  \bibfield  {author} {\bibinfo {author} {\bibfnamefont {S.}~\bibnamefont
  {Tomi\'{c}}}, \bibinfo {author} {\bibfnamefont {L.}~\bibnamefont
  {Bernasconi}}, \bibinfo {author} {\bibfnamefont {B.~G.}\ \bibnamefont
  {Searle}}, \ and\ \bibinfo {author} {\bibfnamefont {N.~M.}\ \bibnamefont
  {Harrison}},\ }\href@noop {} {\bibfield  {journal} {\bibinfo  {journal} {J.
  Phys. Chem. C}\ }\textbf {\bibinfo {volume} {118}},\ \bibinfo {pages} {14478}
  (\bibinfo {year} {2014})}\BibitemShut {NoStop}%
\bibitem [{\citenamefont {Ferrari}\ \emph {et~al.}(2015)\citenamefont
  {Ferrari}, \citenamefont {Orlando},\ and\ \citenamefont
  {R\'{e}rat}}]{Ferrari2015}%
  \BibitemOpen
  \bibfield  {author} {\bibinfo {author} {\bibfnamefont {A.~M.}\ \bibnamefont
  {Ferrari}}, \bibinfo {author} {\bibfnamefont {R.}~\bibnamefont {Orlando}}, \
  and\ \bibinfo {author} {\bibfnamefont {M.}~\bibnamefont {R\'{e}rat}},\
  }\href@noop {} {\bibfield  {journal} {\bibinfo  {journal} {J. Chem. Teho.
  Comp.}\ }\textbf {\bibinfo {volume} {11}},\ \bibinfo {pages} {3245} (\bibinfo
  {year} {2015})}\BibitemShut {NoStop}%
\bibitem [{\citenamefont {Jain}\ \emph {et~al.}(2011)\citenamefont {Jain},
  \citenamefont {Chelikowsky},\ and\ \citenamefont {Louie}}]{Jain2011}%
  \BibitemOpen
  \bibfield  {author} {\bibinfo {author} {\bibfnamefont {M.}~\bibnamefont
  {Jain}}, \bibinfo {author} {\bibfnamefont {J.~R.}\ \bibnamefont
  {Chelikowsky}}, \ and\ \bibinfo {author} {\bibfnamefont {S.~G.}\ \bibnamefont
  {Louie}},\ }\href@noop {} {\bibfield  {journal} {\bibinfo  {journal} {Phys.
  Rev. Lett}\ }\textbf {\bibinfo {volume} {107}},\ \bibinfo {pages} {216806}
  (\bibinfo {year} {2011})}\BibitemShut {NoStop}%
\bibitem [{\citenamefont {Refaely-Abramson}\ \emph {et~al.}(2013)\citenamefont
  {Refaely-Abramson}, \citenamefont {Sharifzadeh}, \citenamefont {Jain},
  \citenamefont {Baer}, \citenamefont {Neaton},\ and\ \citenamefont
  {Kronik}}]{Refaely-Abramson2013}%
  \BibitemOpen
  \bibfield  {author} {\bibinfo {author} {\bibfnamefont {S.}~\bibnamefont
  {Refaely-Abramson}}, \bibinfo {author} {\bibfnamefont {S.}~\bibnamefont
  {Sharifzadeh}}, \bibinfo {author} {\bibfnamefont {M.}~\bibnamefont {Jain}},
  \bibinfo {author} {\bibfnamefont {R.}~\bibnamefont {Baer}}, \bibinfo {author}
  {\bibfnamefont {J.~B.}\ \bibnamefont {Neaton}}, \ and\ \bibinfo {author}
  {\bibfnamefont {L.}~\bibnamefont {Kronik}},\ }\href@noop {} {\bibfield
  {journal} {\bibinfo  {journal} {Phys. Rev. B}\ }\textbf {\bibinfo {volume}
  {88}},\ \bibinfo {pages} {081204(R)} (\bibinfo {year} {2013})}\BibitemShut
  {NoStop}%
\bibitem [{\citenamefont {Yang}\ \emph {et~al.}(2015)\citenamefont {Yang},
  \citenamefont {Sottile},\ and\ \citenamefont {Ullrich}}]{Yang2015}%
  \BibitemOpen
  \bibfield  {author} {\bibinfo {author} {\bibfnamefont {Z.}~\bibnamefont
  {Yang}}, \bibinfo {author} {\bibfnamefont {F.}~\bibnamefont {Sottile}}, \
  and\ \bibinfo {author} {\bibfnamefont {C.~A.}\ \bibnamefont {Ullrich}},\
  }\href@noop {} {\bibfield  {journal} {\bibinfo  {journal} {Phys. Rev. B}\
  }\textbf {\bibinfo {volume} {92}},\ \bibinfo {pages} {035202} (\bibinfo
  {year} {2015})}\BibitemShut {NoStop}%
\bibitem [{\citenamefont {Baer}\ \emph {et~al.}(2010)\citenamefont {Baer},
  \citenamefont {Livshits},\ and\ \citenamefont {Salzner}}]{Baer2010}%
  \BibitemOpen
  \bibfield  {author} {\bibinfo {author} {\bibfnamefont {R.}~\bibnamefont
  {Baer}}, \bibinfo {author} {\bibfnamefont {E.}~\bibnamefont {Livshits}}, \
  and\ \bibinfo {author} {\bibfnamefont {U.}~\bibnamefont {Salzner}},\
  }\href@noop {} {\bibfield  {journal} {\bibinfo  {journal} {Annu. Rev. Phys.
  Chem.}\ }\textbf {\bibinfo {volume} {61}},\ \bibinfo {pages} {85} (\bibinfo
  {year} {2010})}\BibitemShut {NoStop}%
\bibitem [{\citenamefont {Stein}\ \emph {et~al.}(2010)\citenamefont {Stein},
  \citenamefont {Eisenberg}, \citenamefont {Kronik},\ and\ \citenamefont
  {Baer}}]{Stein2010}%
  \BibitemOpen
  \bibfield  {author} {\bibinfo {author} {\bibfnamefont {T.}~\bibnamefont
  {Stein}}, \bibinfo {author} {\bibfnamefont {H.}~\bibnamefont {Eisenberg}},
  \bibinfo {author} {\bibfnamefont {L.}~\bibnamefont {Kronik}}, \ and\ \bibinfo
  {author} {\bibfnamefont {R.}~\bibnamefont {Baer}},\ }\href@noop {} {\bibfield
   {journal} {\bibinfo  {journal} {Phys. Rev. Lett.}\ }\textbf {\bibinfo
  {volume} {105}},\ \bibinfo {pages} {266802} (\bibinfo {year}
  {2010})}\BibitemShut {NoStop}%
\bibitem [{\citenamefont {Refaely-Abramson}\ \emph {et~al.}(2011)\citenamefont
  {Refaely-Abramson}, \citenamefont {Baer},\ and\ \citenamefont
  {Kronik}}]{Refaely-Abramson2011}%
  \BibitemOpen
  \bibfield  {author} {\bibinfo {author} {\bibfnamefont {S.}~\bibnamefont
  {Refaely-Abramson}}, \bibinfo {author} {\bibfnamefont {R.}~\bibnamefont
  {Baer}}, \ and\ \bibinfo {author} {\bibfnamefont {L.}~\bibnamefont
  {Kronik}},\ }\href@noop {} {\bibfield  {journal} {\bibinfo  {journal} {Phys.
  Rev. B}\ }\textbf {\bibinfo {volume} {84}},\ \bibinfo {pages} {075144}
  (\bibinfo {year} {2011})}\BibitemShut {NoStop}%
\bibitem [{RSH()}]{RSHpapers}%
  \BibitemOpen
  \href@noop {} {\ }\bibinfo {note} {See, e.g., C. Risko and J-L Br\'{e}das in
  Top. Curr. Chem. 352, 1 (2014); J. Autschbach and M. Srebro, Acc. Chem. Rev.
  47, 2592 (2014) ; C. Faber, P. Boulanger, C. Attaccalite, I. Duchemin and X.
  Blase, Phil. Trans. R. Soc. A 372, 20130271 (2014); H. Phillips, S. Zheng, E.
  Geva, and B. D. Dunietz, Org. Electronics 7, 1509 (2014); M. E. Foster, J. D.
  Azoulay, B. M. Wong, and M. D. Allendorf, Chem. Sci. 5, 2081 (2014); J. V.
  Koppen, M. Hapka, M. M. Szcz²e\'{s}niak, and G. Cha³asi\'{n}ski, J. Chem.
  Phys. 137, 114302 (2012); T. K\"{o}rzd\"{o}rfer and J-L. Br\'{e}das, Acc.
  Chem. Res. 47, 3284 (2014).}\BibitemShut {Stop}%
\bibitem [{\citenamefont {Refaely-Abramson}\ \emph {et~al.}(2012)\citenamefont
  {Refaely-Abramson}, \citenamefont {Sharifzadeh}, \citenamefont {Govind},
  \citenamefont {Autschbach}, \citenamefont {Neaton}, \citenamefont {Baer},\
  and\ \citenamefont {Kronik}}]{Refaely-Abramson2012}%
  \BibitemOpen
  \bibfield  {author} {\bibinfo {author} {\bibfnamefont {S.}~\bibnamefont
  {Refaely-Abramson}}, \bibinfo {author} {\bibfnamefont {S.}~\bibnamefont
  {Sharifzadeh}}, \bibinfo {author} {\bibfnamefont {N.}~\bibnamefont {Govind}},
  \bibinfo {author} {\bibfnamefont {J.}~\bibnamefont {Autschbach}}, \bibinfo
  {author} {\bibfnamefont {J.~B.}\ \bibnamefont {Neaton}}, \bibinfo {author}
  {\bibfnamefont {R.}~\bibnamefont {Baer}}, \ and\ \bibinfo {author}
  {\bibfnamefont {L.}~\bibnamefont {Kronik}},\ }\href@noop {} {\bibfield
  {journal} {\bibinfo  {journal} {Phys. Rev. Lett.}\ }\textbf {\bibinfo
  {volume} {109}},\ \bibinfo {pages} {226405} (\bibinfo {year}
  {2012})}\BibitemShut {NoStop}%
\bibitem [{\citenamefont {Egger}\ \emph {et~al.}(2014)\citenamefont {Egger},
  \citenamefont {Weissman}, \citenamefont {Refaely-Abramson}, \citenamefont
  {Sharifzadeh}, \citenamefont {Dauth}, \citenamefont {Baer}, \citenamefont
  {K\"{u}mmel}, \citenamefont {Neaton}, \citenamefont {Zojer},\ and\
  \citenamefont {Kronik}}]{Egger2014}%
  \BibitemOpen
  \bibfield  {author} {\bibinfo {author} {\bibfnamefont {D.~A.}\ \bibnamefont
  {Egger}}, \bibinfo {author} {\bibfnamefont {S.}~\bibnamefont {Weissman}},
  \bibinfo {author} {\bibfnamefont {S.}~\bibnamefont {Refaely-Abramson}},
  \bibinfo {author} {\bibfnamefont {S.}~\bibnamefont {Sharifzadeh}}, \bibinfo
  {author} {\bibfnamefont {M.}~\bibnamefont {Dauth}}, \bibinfo {author}
  {\bibfnamefont {R.}~\bibnamefont {Baer}}, \bibinfo {author} {\bibfnamefont
  {S.}~\bibnamefont {K\"{u}mmel}}, \bibinfo {author} {\bibfnamefont {J.~B.}\
  \bibnamefont {Neaton}}, \bibinfo {author} {\bibfnamefont {E.}~\bibnamefont
  {Zojer}}, \ and\ \bibinfo {author} {\bibfnamefont {L.}~\bibnamefont
  {Kronik}},\ }\href@noop {} {\bibfield  {journal} {\bibinfo  {journal} {J.
  Chem. Thoery Comput.}\ }\textbf {\bibinfo {volume} {10}},\ \bibinfo {pages}
  {1934} (\bibinfo {year} {2014})}\BibitemShut {NoStop}%
\bibitem [{Ngu()}]{Nguyen2015}%
  \BibitemOpen
  \href@noop {} {\ }\bibinfo {note} {For a related approach, see N. L. Nguyen,
  G. Borghi, A. Ferretti, I. Dabo, and N. Marzari, Phys. Rev. Lett 114, 166405
  (2015); I. Dabo, A. Ferretti, N. Poilvert, Y. Li, N. Marzari, M. Cococcioni,
  Phys. Rev. B 82, 115121(2010).}\BibitemShut {Stop}%
\bibitem [{\citenamefont {L\"{u}ftner}\ \emph {et~al.}(2014)\citenamefont
  {L\"{u}ftner}, \citenamefont {Refaely-Abramson}, \citenamefont {Pachler},
  \citenamefont {Resel}, \citenamefont {Ramsey}, \citenamefont {Kronik},\ and\
  \citenamefont {Puschnig}}]{Lueftner2014}%
  \BibitemOpen
  \bibfield  {author} {\bibinfo {author} {\bibfnamefont {D.}~\bibnamefont
  {L\"{u}ftner}}, \bibinfo {author} {\bibfnamefont {S.}~\bibnamefont
  {Refaely-Abramson}}, \bibinfo {author} {\bibfnamefont {M.}~\bibnamefont
  {Pachler}}, \bibinfo {author} {\bibfnamefont {R.}~\bibnamefont {Resel}},
  \bibinfo {author} {\bibfnamefont {M.~G.}\ \bibnamefont {Ramsey}}, \bibinfo
  {author} {\bibfnamefont {L.}~\bibnamefont {Kronik}}, \ and\ \bibinfo {author}
  {\bibfnamefont {P.}~\bibnamefont {Puschnig}},\ }\href@noop {} {\bibfield
  {journal} {\bibinfo  {journal} {Phys. Rev. B}\ }\textbf {\bibinfo {volume}
  {90}},\ \bibinfo {pages} {075204} (\bibinfo {year} {2014})}\BibitemShut
  {NoStop}%
\bibitem [{\citenamefont {Leininger}\ \emph {et~al.}(1997)\citenamefont
  {Leininger}, \citenamefont {Stoll}, \citenamefont {Werner},\ and\
  \citenamefont {Savin}}]{Leininger1997}%
  \BibitemOpen
  \bibfield  {author} {\bibinfo {author} {\bibfnamefont {T.}~\bibnamefont
  {Leininger}}, \bibinfo {author} {\bibfnamefont {H.}~\bibnamefont {Stoll}},
  \bibinfo {author} {\bibfnamefont {H.~J.}\ \bibnamefont {Werner}}, \ and\
  \bibinfo {author} {\bibfnamefont {A.}~\bibnamefont {Savin}},\ }\href@noop {}
  {\bibfield  {journal} {\bibinfo  {journal} {Chem. Phys. Lett.}\ }\textbf
  {\bibinfo {volume} {275}},\ \bibinfo {pages} {151} (\bibinfo {year}
  {1997})}\BibitemShut {NoStop}%
\bibitem [{\citenamefont {Yanai}\ \emph {et~al.}(2004)\citenamefont {Yanai},
  \citenamefont {Tew},\ and\ \citenamefont {Handy}}]{Yanai2004}%
  \BibitemOpen
  \bibfield  {author} {\bibinfo {author} {\bibfnamefont {T.}~\bibnamefont
  {Yanai}}, \bibinfo {author} {\bibfnamefont {D.~P.}\ \bibnamefont {Tew}}, \
  and\ \bibinfo {author} {\bibfnamefont {N.~C.}\ \bibnamefont {Handy}},\
  }\href@noop {} {\bibfield  {journal} {\bibinfo  {journal} {Chem. Phys.
  Lett.}\ }\textbf {\bibinfo {volume} {393}},\ \bibinfo {pages} {51} (\bibinfo
  {year} {2004})}\BibitemShut {NoStop}%
\bibitem [{\citenamefont {Tretiak}\ and\ \citenamefont
  {Chernyak}(2003)}]{Tretiak2003}%
  \BibitemOpen
  \bibfield  {author} {\bibinfo {author} {\bibfnamefont {S.}~\bibnamefont
  {Tretiak}}\ and\ \bibinfo {author} {\bibfnamefont {V.}~\bibnamefont
  {Chernyak}},\ }\href@noop {} {\bibfield  {journal} {\bibinfo  {journal} {J.
  Chem. Phys.}\ }\textbf {\bibinfo {volume} {119}},\ \bibinfo {pages} {8809}
  (\bibinfo {year} {2003})}\BibitemShut {NoStop}%
\bibitem [{\citenamefont {Vl\u{c}ek}\ \emph {et~al.}(2014)\citenamefont
  {Vl\u{c}ek}, \citenamefont {Eisenberg}, \citenamefont {Steinle-Neumann},
  \citenamefont {Kronik},\ and\ \citenamefont {Baer}}]{Vlcek2015}%
  \BibitemOpen
  \bibfield  {author} {\bibinfo {author} {\bibfnamefont {V.}~\bibnamefont
  {Vl\u{c}ek}}, \bibinfo {author} {\bibfnamefont {H.~R.}\ \bibnamefont
  {Eisenberg}}, \bibinfo {author} {\bibfnamefont {G.}~\bibnamefont
  {Steinle-Neumann}}, \bibinfo {author} {\bibfnamefont {L.}~\bibnamefont
  {Kronik}}, \ and\ \bibinfo {author} {\bibfnamefont {R.}~\bibnamefont
  {Baer}},\ }\href@noop {} {\bibfield  {journal} {\bibinfo  {journal} {J. Chem.
  Phys.}\ }\textbf {\bibinfo {volume} {142}},\ \bibinfo {pages} {034107}
  (\bibinfo {year} {2014})}\BibitemShut {NoStop}%
\bibitem [{\citenamefont {Ihm}\ \emph {et~al.}(1979)\citenamefont {Ihm},
  \citenamefont {Zunger},\ and\ \citenamefont {Cohen}}]{Ihm1979}%
  \BibitemOpen
  \bibfield  {author} {\bibinfo {author} {\bibfnamefont {J.}~\bibnamefont
  {Ihm}}, \bibinfo {author} {\bibfnamefont {A.}~\bibnamefont {Zunger}}, \ and\
  \bibinfo {author} {\bibfnamefont {M.~L.}\ \bibnamefont {Cohen}},\ }\href@noop
  {} {\bibfield  {journal} {\bibinfo  {journal} {J. Phys. C: Solid State
  Phys.}\ }\textbf {\bibinfo {volume} {12}},\ \bibinfo {pages} {4409} (\bibinfo
  {year} {1979})}\BibitemShut {NoStop}%
\bibitem [{\citenamefont {Deslippe}\ \emph {et~al.}(2012)\citenamefont
  {Deslippe}, \citenamefont {Samsonidze}, \citenamefont {Strubbe},
  \citenamefont {Jain}, \citenamefont {Cohen},\ and\ \citenamefont
  {Louie}}]{Deslippe2012}%
  \BibitemOpen
  \bibfield  {author} {\bibinfo {author} {\bibfnamefont {J.}~\bibnamefont
  {Deslippe}}, \bibinfo {author} {\bibfnamefont {G.}~\bibnamefont
  {Samsonidze}}, \bibinfo {author} {\bibfnamefont {D.~A.}\ \bibnamefont
  {Strubbe}}, \bibinfo {author} {\bibfnamefont {M.}~\bibnamefont {Jain}},
  \bibinfo {author} {\bibfnamefont {M.~L.}\ \bibnamefont {Cohen}}, \ and\
  \bibinfo {author} {\bibfnamefont {S.~G.}\ \bibnamefont {Louie}},\ }\href@noop
  {} {\bibfield  {journal} {\bibinfo  {journal} {Comput. Phys. Commun.}\
  }\textbf {\bibinfo {volume} {183}},\ \bibinfo {pages} {1269} (\bibinfo {year}
  {2012})}\BibitemShut {NoStop}%
\bibitem [{\citenamefont {Chelikowsky}\ and\ \citenamefont
  {Cohen}(1976)}]{CC1976}%
  \BibitemOpen
  \bibfield  {author} {\bibinfo {author} {\bibfnamefont {J.~R.}\ \bibnamefont
  {Chelikowsky}}\ and\ \bibinfo {author} {\bibfnamefont {M.~L.}\ \bibnamefont
  {Cohen}},\ }\href@noop {} {\bibfield  {journal} {\bibinfo  {journal} {Phys.
  Rev. B}\ }\textbf {\bibinfo {volume} {14}},\ \bibinfo {pages} {556} (\bibinfo
  {year} {1976})}\BibitemShut {NoStop}%
\bibitem [{\citenamefont {Lautenschlager}\ \emph {et~al.}(1987)\citenamefont
  {Lautenschlager}, \citenamefont {Garriga}, \citenamefont {Vina},\ and\
  \citenamefont {Cardona}}]{Lautenschlager1987}%
  \BibitemOpen
  \bibfield  {author} {\bibinfo {author} {\bibfnamefont {P.}~\bibnamefont
  {Lautenschlager}}, \bibinfo {author} {\bibfnamefont {M.}~\bibnamefont
  {Garriga}}, \bibinfo {author} {\bibfnamefont {L.}~\bibnamefont {Vina}}, \
  and\ \bibinfo {author} {\bibfnamefont {M.}~\bibnamefont {Cardona}},\
  }\href@noop {} {\bibfield  {journal} {\bibinfo  {journal} {Phys. Rev. B}\
  }\textbf {\bibinfo {volume} {36}},\ \bibinfo {pages} {4821} (\bibinfo {year}
  {1987})}\BibitemShut {NoStop}%
\bibitem [{\citenamefont {Roessler}\ and\ \citenamefont
  {Walker}(1967)}]{Roessler1967}%
  \BibitemOpen
  \bibfield  {author} {\bibinfo {author} {\bibfnamefont {D.~M.}\ \bibnamefont
  {Roessler}}\ and\ \bibinfo {author} {\bibfnamefont {W.~C.}\ \bibnamefont
  {Walker}},\ }\href@noop {} {\bibfield  {journal} {\bibinfo  {journal} {J.
  Opt. Soc. Am}\ }\textbf {\bibinfo {volume} {57}},\ \bibinfo {pages} {835}
  (\bibinfo {year} {1967})}\BibitemShut {NoStop}%
\bibitem [{\citenamefont {Tiago}\ \emph {et~al.}(2003)\citenamefont {Tiago},
  \citenamefont {Northrup},\ and\ \citenamefont {Louie}}]{Tiago2003}%
  \BibitemOpen
  \bibfield  {author} {\bibinfo {author} {\bibfnamefont {M.~L.}\ \bibnamefont
  {Tiago}}, \bibinfo {author} {\bibfnamefont {J.~E.}\ \bibnamefont {Northrup}},
  \ and\ \bibinfo {author} {\bibfnamefont {S.~G.}\ \bibnamefont {Louie}},\
  }\href@noop {} {\bibfield  {journal} {\bibinfo  {journal} {Phys. Rev. B}\
  }\textbf {\bibinfo {volume} {67}},\ \bibinfo {pages} {115212} (\bibinfo
  {year} {2003})}\BibitemShut {NoStop}%
\bibitem [{\citenamefont {Cudazzo}\ \emph {et~al.}(2012)\citenamefont
  {Cudazzo}, \citenamefont {Gatti},\ and\ \citenamefont {Rubio}}]{Cudazzo2012}%
  \BibitemOpen
  \bibfield  {author} {\bibinfo {author} {\bibfnamefont {P.}~\bibnamefont
  {Cudazzo}}, \bibinfo {author} {\bibfnamefont {M.}~\bibnamefont {Gatti}}, \
  and\ \bibinfo {author} {\bibfnamefont {A.}~\bibnamefont {Rubio}},\
  }\href@noop {} {\bibfield  {journal} {\bibinfo  {journal} {Phys. Rev. B}\
  }\textbf {\bibinfo {volume} {86}},\ \bibinfo {pages} {195307} (\bibinfo
  {year} {2012})}\BibitemShut {NoStop}%
\bibitem [{\citenamefont {Sharifzadeh}\ \emph {et~al.}(2013)\citenamefont
  {Sharifzadeh}, \citenamefont {Darancet}, \citenamefont {Kronik},\ and\
  \citenamefont {Neaton}}]{Sharifzadeh2013}%
  \BibitemOpen
  \bibfield  {author} {\bibinfo {author} {\bibfnamefont {S.}~\bibnamefont
  {Sharifzadeh}}, \bibinfo {author} {\bibfnamefont {P.}~\bibnamefont
  {Darancet}}, \bibinfo {author} {\bibfnamefont {L.}~\bibnamefont {Kronik}}, \
  and\ \bibinfo {author} {\bibfnamefont {J.~B.}\ \bibnamefont {Neaton}},\
  }\href@noop {} {\bibfield  {journal} {\bibinfo  {journal} {J. Phys. Chem.
  Lett.}\ }\textbf {\bibinfo {volume} {4}},\ \bibinfo {pages} {2197} (\bibinfo
  {year} {2013})}\BibitemShut {NoStop}%
\bibitem [{\citenamefont {Sharifzadeh}\ \emph {et~al.}(2012)\citenamefont
  {Sharifzadeh}, \citenamefont {Biller}, \citenamefont {Kronik},\ and\
  \citenamefont {Neaton}}]{Sharifzadeh2012-2}%
  \BibitemOpen
  \bibfield  {author} {\bibinfo {author} {\bibfnamefont {S.}~\bibnamefont
  {Sharifzadeh}}, \bibinfo {author} {\bibfnamefont {A.}~\bibnamefont {Biller}},
  \bibinfo {author} {\bibfnamefont {L.}~\bibnamefont {Kronik}}, \ and\ \bibinfo
  {author} {\bibfnamefont {J.~B.}\ \bibnamefont {Neaton}},\ }\href@noop {}
  {\bibfield  {journal} {\bibinfo  {journal} {Phys. Rev. B}\ }\textbf {\bibinfo
  {volume} {85}},\ \bibinfo {pages} {125307} (\bibinfo {year}
  {2012})}\BibitemShut {NoStop}%
\bibitem [{\citenamefont {Rohrdanz}\ \emph {et~al.}(2009)\citenamefont
  {Rohrdanz}, \citenamefont {Martines},\ and\ \citenamefont
  {Herbert}}]{Rohrdanz2009}%
  \BibitemOpen
  \bibfield  {author} {\bibinfo {author} {\bibfnamefont {M.~A.}\ \bibnamefont
  {Rohrdanz}}, \bibinfo {author} {\bibfnamefont {K.~M.}\ \bibnamefont
  {Martines}}, \ and\ \bibinfo {author} {\bibfnamefont {J.~M.}\ \bibnamefont
  {Herbert}},\ }\href@noop {} {\bibfield  {journal} {\bibinfo  {journal} {J.
  Chem. Phys.}\ }\textbf {\bibinfo {volume} {130}},\ \bibinfo {pages} {054112}
  (\bibinfo {year} {2009})}\BibitemShut {NoStop}%
\end{thebibliography}%

\end{document}